\documentclass[12pt]{article}

\usepackage{graphicx}
\usepackage{amssymb}

\begin{document}

\input{epsf.sty}

\begin{titlepage}

\begin{flushright}
\end{flushright}
\vskip 2.5cm

\begin{center}
{\Large \bf Finite Duration and Energy Effects in\\
Lorentz-Violating Vacuum Cerenkov Radiation}
\end{center}

\vspace{1ex}

\begin{center}
{\large Brett Altschul\footnote{{\tt baltschu@physics.sc.edu}}}

\vspace{5mm}
{\sl Department of Physics and Astronomy} \\
{\sl University of South Carolina} \\
{\sl Columbia, SC 29208 USA} \\

\end{center}

\vspace{2.5ex}

\medskip

\centerline {\bf Abstract}

\bigskip

Vacuum Cerenkov radiation is possible in certain Lorentz-violating quantum
field theories, when very energetic charges move faster than the phase speed of
light. In the presence of a CPT-even, Lorentz-violating modification
of the photon sector,
the character of the Cerenkov process is controlled by the high-frequency
behavior of the radiation spectrum. The development of the Cerenkov process
can be markedly different,
depending on whether the only limits on the emission of very energetic photons come
from energy-momentum conservation or whether there are additional effects that
cut off the spectrum at high frequencies.
Moreover, since the high-frequency cutoff determines the
total rate at which an emitting charge loses energy, it also controls all aspects of
the emission that are related to the process's finite duration.

\bigskip

\end{titlepage}

\newpage

\section{Introduction}

In roughly the last decade, there has been a great surge in interest in the
possibility that Lorentz invariance may not be exact. If Lorentz violation were
discovered experimentally, it would be a discovery of tremendous significance and
would mean that there existed qualitatively new physics beyond
general relativity and the standard model. However, despite many precision tests,
there is thus far no evidence that relativity needs any modification.

Precision tests of Lorentz invariance are nothing new, but the field of Lorentz
violation has changed substantially in recent years. For a long time, most tests of
relativity were designed to search for {\em ad hoc} modifications of standard
relativistic physics. That changed with the development of a systematic effective
field theory approach. The standard model extension (SME) is an effective field
theory that incorporates known physics and also the possibility of Lorentz
violation~\cite{ref-kost1,ref-kost2}. The violations enter through Lorentz
noninvariant operators in the Lagrangian, parameterized by coefficient tensors with
Lorentz indices. If Lorentz symmetry is broken spontaneously, these coefficients are
the vacuum expectation values of tensor operators, selecting out preferred
directions in spacetime. Some of these operators violate, in addition to
Lorentz invariance, CPT invariance.

There are many ways that Lorentz symmetry can be violated in the SME. If
nonrenormalizable terms are included in the Lagrangian, the number of coefficients
characterizing the theory is infinite. The minimal SME contains only local,
gauge invariant operators of dimension four or less that can be constructed out of
standard model fields. The number of coefficients is still very large, but in most
situations, only a relatively modest subset of them will affect a particular
observable. For example, in many cases, only the Lorentz-violating coefficients for
protons, neutrons, electrons, and photons come into play. These are the species
we observe in low-energy physics experiments, and Lorentz violations in these
sectors are fairly well bounded (whereas this is not so much the case for more
exotic particles and fields).

Some effects which are absolutely forbidden in Lorentz invariant theories can
occur readily in the SME. When the action is no longer invariant under Lorentz
boosts, it is possible for different particles to have different maximum velocities.
Specifically, it is possible for some particles to travel faster than the
phase velocity
of light (which is not necessarily energy independent). When charged particles
move this fast, they must emit vacuum Cerenkov radiation.
This kind of radiation is a unique signature of Lorentz violation, and it has
already received a fair amount of attention~\cite{ref-lehnert1,ref-lehnert2,
ref-kaufhold1,ref-jacobson3,ref-altschul9,ref-altschul12,ref-kaufhold2}.
However, vacuum
Cerenkov radiation is by no means completely understood. In particular, it is not
entirely clear what kind of role new physics entering at large energy scales will
play in the process. We shall address that particular question in this paper.
The expression for the total power radiated off by a superluminal charged particle
is dominated by the ultraviolet end of the frequency spectrum. This means that
anything dependent on this total power will depend on the spectrum's high-energy
cutoff. The total emission rate determines all the properties of the Cerenkov
process that are tied to its finite duration.

We shall therefore focus on how finite energy and finite duration effects play out
in the vacuum Cerenkov process. The paper is organized as follows. In the remainder
of the introduction, we shall consider the particular CPT-even model that will be
discussed in the rest of the paper. We shall then examine the effects of the one
cutoff for the emission spectrum that is guaranteed to exist---the cutoff due to
energy-momentum conservation---in section~\ref{sec-recoil}. In
section~\ref{sec-cutoffs}, we examine the competing effects of other possible
cutoffs, including one whose existence is strongly suggested by naturalness
considerations. Then we turn in section~\ref{sec-diff} to a study of diffraction
in the Cerenkov process; this topic is interesting in itself and also draws
together many of the results from earlier in the paper. We conclude in
section~\ref{sec-concl} with a narrative describing how the vacuum Cerenkov process
evolves over time and some additional remarks.

The least constrained operators in the photon sector of the minimal SME are part of
the tensor $k_{F}^{\mu\nu\rho\sigma}$ appearing in the electromagnetic Lagrange
density
\begin{equation}
{\cal L}_{F}=-\frac{1}{4}F^{\mu\nu}F_{\mu\nu}
-\frac{1}{4}k_{F}^{\mu\nu\rho\sigma}F_{\mu\nu}F_{\rho\sigma}.
\end{equation}
There are nineteen independent parameters contained in the CPT-even
$k_{F}^{\mu\nu\rho\sigma}$. Ten of them are associated with photon birefringence
and have been very strongly constrained with cosmological
measurements~\cite{ref-kost11,ref-kost21}. (All the CPT-odd parameters are
likewise very tightly bounded~\cite{ref-carroll1,ref-carroll2}.) The terms that do
not lead to birefringence form a two-index traceless symmetric tensor
$\tilde{k}^{\mu\nu}$, and if
all the birefringent terms are zero,
\begin{equation}
\label{eq-kF}
k_{F}^{\mu\nu\rho\sigma}=\frac{1}{2}\left(g^{\mu\rho}\tilde{k}^{\nu\sigma}
-g^{\mu\sigma}\tilde{k}^{\nu\rho}
-g^{\nu\rho}\tilde{k}^{\mu\sigma}
+g^{\nu\sigma}\tilde{k}^{\mu\rho}\right).
\end{equation}
Most of the results in this paper can be generalized to cover the case of the most
general $k_{F}^{\mu\nu\rho\sigma}$. We must simply split up the two polarizations,
which propagate at different rates. Calculations of the Cerenkov spectrum in the
presence of this birefringence is discussed in detail in~\cite{ref-altschul12}.
However, we shall restrict our explicit calculations here to the
$\tilde{k}^{\mu\nu}$ only case. We shall work to leading order in
$\tilde{k}^{\mu\nu}$, because Lorentz violation is supposed to be a small effect,
and any higher order corrections must be miniscule.

For simplicity, we shall also
not consider
directly any modifications to the charged matter, which will generally affect the
relationships between charged particles' momenta and velocities. Perhaps the
simplest form of Lorenz violation for a fermion is given by
\begin{equation}
{\cal L}_{\psi}=\bar{\psi}[(\gamma^{\mu}+c^{\nu\mu}\gamma_{\nu})
i\partial_{\mu}-m]\psi.
\end{equation}
Radiative corrections mix $\tilde{k}^{\mu\nu}$ and the $c^{\nu\mu}$ terms for charged
species, so we do not expect the matter sector in the presence of
$\tilde{k}^{\mu\nu}$ to be truly conventional.
However, for the purposes of determining the vacuum Cerenkov radiation, we may
assume that the $c^{\nu\mu}$ relevant to the moving charges vanishes, so long as
we henceforth take the effective $\tilde{k}^{\mu\nu}$ to be $\tilde{k}_{0}^{\mu\nu}
-2c^{\nu\mu}$, where $\tilde{k}_{0}^{\mu\nu}$ is the true Lorentz-violating
parameter appearing in the photon Lagrangian~\cite{ref-kost17}.

We shall be studying what happens to a charged particle moving in a direction
$\hat{v}$ with a speed $v$ very close to 1. The phase speed at which light propagates
in this direction is $1-\frac{1}{2}\left[\tilde{k}_{jk}\hat{v}_{j}\hat{v}_{k}+
2\tilde{k}_{0j}\hat{v}_{j}+\tilde{k}_{00}\right]$. If  $v$ is
greater than this, Cerenkov radiation will occur. Since the phase speed of the
radiation can only deviate very slightly from 1, the Cerenkov cone will be very
narrow. All the radiation is beamed into a narrow pencil of angles around $\hat{v}$,
and so the direction dependence of the phase speed can be ignored. (However, if we
did take into account the fact that the
emitted photons do not travel in precisely the same direction $\hat{v}$ as the
charge, there would be higher order corrections that would deform the Cerenkov cone
so that it would no longer be right angled or circular.)

Since no directions other than $\hat{v}$ are involved in the Cerenkov process,
the leading order effects are generally identical to what one would see if the
charge were moving with a speed $v$ in an isotropic medium with index of
refraction
\begin{equation}
n=\left\{1-\frac{1}{2}\left[\tilde{k}_{jk}\hat{v}_{j}\hat{v}_{k}+
2\tilde{k}_{0j}\hat{v}_{j}+\tilde{k}_{00}\right]\right\}^{-1}.
\end{equation}
If $n\leq1$, Cerenkov radiation is obviously impossible. If $n>1$, there is
radiation if the charge's energy exceeds the threshold energy
\begin{equation}
\label{eq-threshold}
E_{T}=\frac{m}{\sqrt{\tilde{k}_{jk}\hat{v}_{j}\hat{v}_{k}+
2\tilde{k}_{0j}\hat{v}_{j}+\tilde{k}_{00}}}.
\end{equation}
To make contact with the usual expressions describing Cerenkov radiation, which
were derived for the case of radiation in a medium, we shall make frequent
use of the effective refractive index $n$.

With $n$ and $v$ alone, it is already possible to calculate such quantities as
the Cerenkov angle and the power spectrum for a steady state Cerenkov process.
However, this does not capture all the relevant physics; other effects are also quite
important. The steady state analysis neglects
recoil corrections, which are related to the corpuscular nature of light. One
obvious role that the backreaction on a radiating charge must play is as an
ultraviolet regulator for the total power emitted. A charge cannot radiate away more
energy than it possesses. With the backreaction taken into account, the radiation
must cease after a finite time, and when the radiating track length is finite,
diffraction can play an interesting role. Moreover,
new physics that is important only at high energies might also come into play.

\section{Recoil Corrections}
\label{sec-recoil}

Before we consider the impact of any new physics, we should look at how the well
understood effects of energy and momentum conservation affect the Lorentz-violating
Cerenkov process. Obviously, a charge cannot emit photons with
arbitrarily high frequencies; the energy is simply not available. The details of
how recoil corrects the Cerenkov spectrum are comparatively simple, and the
relevant calculations generally mirror those relevant to Cerenkov radiation in
media. We can carry over the standard results using our prescription for $n$ and
making any additional approximations that are appropriate.

There is a well known result for the maximum frequency present in the Cerenkov
spectrum emitted by a charge moving in a perfect, nondispersive
dielectric~\cite{ref-jelley}.
This frequency is determined by energy-momentum conservation during the emission of a
single photon. The maximum frequency for a charge with energy $E$ and speed $v$ is
\begin{equation}
\omega_{m}=2E\frac{vn-1}{n^{2}-1},
\end{equation}
where $n$ is the index of refraction. We shall recast this expression in a
more useful form. At ultrarelativistic energies, where
the charge's energy is approximately $E=m/\sqrt{2(1-v)}$, so that
\begin{equation}
\label{eq-v}
v=1-\frac{m^{2}}{2E^{2}},
\end{equation}
this reduces to
\begin{equation}
\omega_{m}=\frac{2E}{n+1}-\frac{m^{2}}{E}\frac{n}{n^{2}-1}.
\end{equation}
In this regime, 
The threshold $E_{T}$ for low-frequency Cerenkov emission is the energy at which
$v$ is equal to the
speed of light $n^{-1}$ in the medium. If $n$ is close to 1, so $1-n^{-1}\approx
n-1\approx m^{2}/2E_{T}^{2}$, then the expression for $\omega_{m}$ becomes
\begin{equation}
\omega_{m}=\frac{E^{2}-E_{T}^{2}}{E}.
\end{equation}

Importantly, $\omega_{m}=(E-E_{T})\left(\frac{E+E_{T}}{E}\right)$ is always
greater than $E-E_{T}$. So whenever there is Cerenkov emission, there is a finite
probability per unit time of emitting a photon near the upper end of the allowed
frequency spectrum, so that the charge drops below the threshold energy, and further
emission is impossible. It is the rate of emission of $\omega>E-E_{T}$ photons that
determines how long the Cerenkov process will last.

A more general version of the recoil analysis gives the change in the Cererkov
angle $\theta_{C}$ due to recoil effects. The modified expression
is~\cite{ref-jelley}
\begin{equation}
\cos\theta_{C}=\frac{1}{vn}\left[1+\frac{\omega}{2E}\left(n^{2}-1\right)\right],
\end{equation}
and $\omega_{m}$ is the frequency at which $\theta_{C}$ shrinks to zero, cutting off
the emission. To leading order in $n-1$, $\theta_{C}$ is given by
\begin{equation}
\label{eq-thetaC2}
\sin^{2}\theta_{C}=2\left[1-\frac{1}{v}+\left(1-\frac{\omega}{E}\right)\frac{n-1}
{v}\right],
\end{equation}
and at high energies, $1-v^{-1}\approx v-1$ is given by (\ref{eq-v}).

As a lowest-order approximation for the emitted power that accounts for recoil
corrections, we may simply use the recoil-corrected value of the Cerenkov angle
in the power spectrum formula, $P(\omega)=\frac{e^{2}}{4\pi}\sin^{2}\theta_{C}
\omega$.
This is essentially a phase space estimate, using a matrix element for the
emission process that does not include recoil corrections but including the full
effects of the recoil in the kinematics.
In this approximation, and to leading order in $n-1$ (and hence leading order in
the Lorentz violation), the rate of photon emission per unit frequency is
\begin{equation}
\Gamma(\omega)=\frac{P(\omega)}{\omega}=
\frac{e^{2}m^{2}}{4\pi}\left[E_{T}^{-2}\left(1-\frac{\omega}{E}
\right)-E^{-2}\right].
\end{equation}
If there is no cutoff other than that provided by the backreaction and
energy-momentum conservation, the instantaneous rate of emission for all photons with
energies greater than $E-E_{T}$ is
\begin{equation}
\label{eq-Gamma}
\Gamma\equiv
\int_{E-E_{T}}^{\omega_{m}}d\omega\,\Gamma(\omega)=\frac{e^{2}m^{2}}{8\pi}
\frac{(E-E_{T})^{2}}{E^{3}}.
\end{equation}
This is the instantaneous decay constant for the process in which the charge emits a
single high-energy photon, drops below the Cerenkov threshold, and consequently stops
emitting.

The rate $\Gamma$ is small when the energy $E$ is only slightly above the
threshold, and it increases to a maximum
value of $\Gamma=\frac{e^{2}m^{2}}{54\pi E_{T}}$ at $E=3E_{T}$, so the
probability per unit time of the Cerenkov process coming to a sudden halt never
exceeds $\sim 10^{-4}\,m\left|\tilde{k}\right|^{1/2}$.
At the highest energies, the
rate behaves as $\Gamma\approx\frac{e^{2}m^{2}}{8\pi E}$.

$\Gamma$ represents the rate for one kind of energy loss. Energy is also lost
through the emission of lower-energy photons with $\omega<E-E_{T}$. The emission of
one of these photons will not lower the energy to
below $E_{T}$, and so the charge will
continue to radiate afterwards. This makes it reasonable to approximate the energy
losses from $\omega<E-E_{T}$ photons as a continuous process, radiating power at a
rate
\begin{equation}
P_{<}\equiv
\int_{0}^{E-E_{T}}d\omega\,P(\omega)=\frac{e^{2}m^{2}}{24\pi}\frac{(E-E_{T})^{3}
(E+3E_{T})}{E^{2}E_{T}^{2}},
\end{equation}
$P_{<}$ increases as $E^{2}$ at large energies, when $E\gg E_{T}$.
In this regime, the time scale for the charge to lose a substantial fraction of its
energy by emitting lower-energy, $\omega<E-E_{T}$, photons is
\begin{equation}
\frac{E}{P_{<}}\approx\frac{24\pi E_{T}^{2}}{e^{2}m^{2}E},
\end{equation}
which
is a much shorter time scale than $\Gamma^{-1}$. At high energies, the continuous
emission of
lower-energy photons is more important than the possibility of a single high-energy
event that drops the particle below threshold. However, when $E$ is only slightly
greater than $E_{T}$, the characteristic scales $(E-E_{T})/P_{<}$ and $\Gamma^{-1}$
for the two types of losses are comparable.

\begin{figure}[t]
\epsfxsize=3in
\begin{center}
\leavevmode
\epsfbox[300 350 550 600]{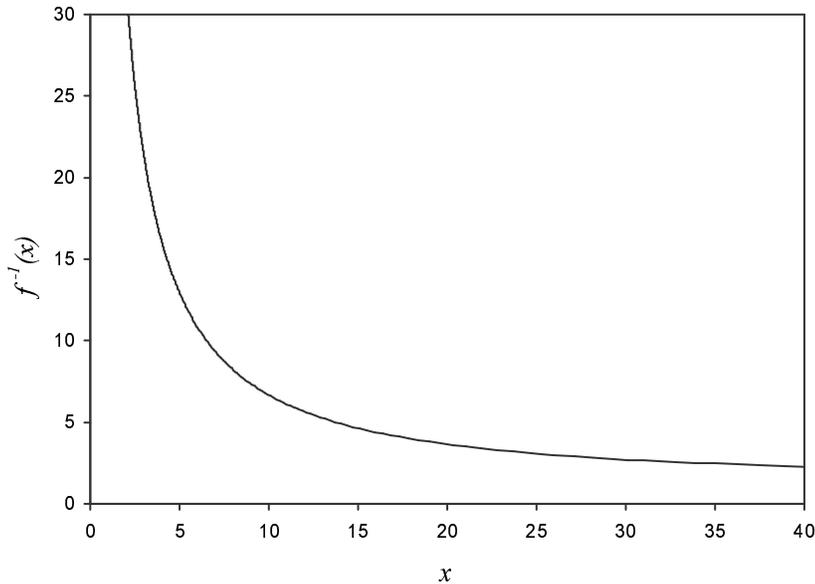}
\caption{The function $f^{-1}(x)$, showing its characteristic behavior both at large
and small $x$.
\label{fig-f-1}}
\end{center}
\end{figure}

We can combine $\Gamma$ and $P_{<}$ to find the time dependence of $\Gamma$.
To do this, we approximate the energy loss coming from the lower-frequency
radiation as deterministic, neglecting the decomposition of the emission into
photons. We can then solve for the energy, given that no photon with an energy
above $E-E_{T}$ is emitted before a time $t$, by solving $\dot{E}=-P_{<}$. The
solution is elementary---
\begin{equation}
E=E_{T}f^{-1}\left[\frac{8e^{2}m^{2}t}{3\pi E_{T}}+f\left(\frac{E_{0}}{E_{T}}
\right)\right],
\end{equation}
where $E_{0}$ is the charge's energy at $t=0$ and $f^{-1}$ is the inverse of the
function
\begin{equation}
f(x)=\frac{4(7x-5)}{(x-1)^{2}}+9\log\frac{x+3}{x-1}.
\end{equation}
Combined with (\ref{eq-Gamma}), this gives the time dependence of $\Gamma$.
The function $f^{-1}(x)$ is plotted in figure~\ref{fig-f-1}.
As $x\rightarrow0$,
$f^{-1}(x)\approx28/x$, and this governs the behavior of the energy
at small times and when $E_{0}$ is large. For large values of $x$,
$f^{-1}(x)\approx1+2\sqrt{2/x}$, indicating a more gradual rate of energy loss as
the energy drops close to $E_{T}$.

\section{Ultraviolet Cutoffs}
\label{sec-cutoffs}

Thus far, the only ultraviolet cutoff for the radiation that we have considered is
$\omega_{m}$, whose existence is guaranteed by energy and momentum conservation.
However, this is not necessarily the only cutoff that might affect the theory or
even the most significant one.
The most important cutoff will be whatever one lies
the lowest in energy.
In this section, we shall look at possibly relevant cutoffs, considering them
separately to see how they compare.

The energy conservation cutoff is $\omega_{m}$. Just above threshold,
when $E-E_{T}\ll E_{T}$, this cutoff may be quite small.  Obviously, for sufficiently
small $E-E_{T}$, this must be the most
relevant cutoff. What happens at greater energies is less clear.
Beyond the $E-E_{T}\ll E_{T}$ regime,
we have $\omega_{m}\approx E$, the cutoff growing linearly
with the charge's energy scale.
From (\ref{eq-thetaC2}), it is evident that the Cerenkov angle---which
governs the emission rate---is little modified by the cutoff except for photon
frequencies
comparable to $E$. The scale $E_{T}$, when it is comparable to or lower than $E$,
does not play any role in
determining the frequency cutoff; nor does $m$. $E$ provides the only scale involved.
On dimensional grounds alone, the rate of energy loss must then be proportional to
$E^{2}$, which indeed it is; the rate at which the charge loses energy smoothly is
just $P_{<}$, which has the required energy dependence.

The electromagnetic sector, including $\tilde{k}^{\mu\nu}$, is invariant under
dilation;
it contains no preferred scale. However, in any theory with massive charged
particles, the scale invariance is broken. There are Lorentz-violating operators
in the charged fermion sector parameterized by coefficients $c^{\nu\mu}$, which mix
with the $\tilde{k}^{\mu\nu}$ operators under renormalization~\cite{ref-kost4}.
In a natural theory,
the $c^{\nu\mu}$ coefficients cannot be smaller than the $\tilde{k}^{\mu\nu}$
[except
possibly by a factor of ${\cal O}(\alpha)$]. What is important about the existence
of the $c^{\nu\mu}$ terms is that they introduce another important scale into the
theory beyond the fermion mass scale. The fermion sector will begin to have problems
with stability or causality when particles reach momenta
$\sim m|c|^{-1/2}$~\cite{ref-kost3}.

What happens at this scale can be understood as follows. The maximum achievable
velocity (MAV) for a species of fermions depends on its $c^{\nu\mu}$. If
the coefficients are such that the MAV in a given direction exceeds 1, then there
are obviously causality problems. Particles with large momenta along the relevant
direction in one frame will be able to travel superluminally, which means backwards
in time as measured in a different observer frame. The stability problems occur if
the MAV is less than 1. Then there are on-shell particle states with spacelike
momenta, and in sufficiently boosted frames, these states will have negative
energies, destabilizing the vacuum. The momentum scale at which either of these
problems fist becomes evident can readily be seen to be $m|c|^{-1/2}$.
Some of the best bounds on electron Lorentz violation actually come from
constraining the deviation of the electron MAV from 1, using, among other
techniques, the observed absence of vacuum Cerenkov radiation in the spectra of
energetic astrophysical
sources~\cite{ref-jacobson1,ref-altschul6,ref-altschul11,ref-altschul13}.

If new physics intervenes at the scale $\Lambda_{c}\sim m|c|^{-1/2}$ to
preserve some form of
causality, we expect the new interactions to deform the effective energy-momentum
relation in such a way as to counteract the effects of $c^{\nu\mu}$. The simplest way
to do
this would be with higher-dimension, energy-dependent operators that keep the
dispersion relations from going outside the null cone at high energies. This is not
the only possibility, however. What is important is that it is natural (although one
cannot say required) that there be new physics entering at energies
$\Lambda_{\tilde{k}}\sim m\left|\tilde{k}\right|^{-1/2}$ which will cut off the
Cerenkov radiation, possibly by restoring the photon dispersion relation to its
conventional $\omega=\left|\vec{k}\right|$ form at higher momenta.

The mass scale
$m$ appearing in $\Lambda_{\tilde{k}}$ represents the mass of the lightest charged
particle (physically, the electron). In principle, each species has its own
coefficients $c^{\nu\mu}$ (which actually do not need to be diagonal in flavor
space), but naturalness dictates that all the $c^{\nu\mu}$ should be comparable in
size. The scales at which causality problems occur are not the same for the various
species in the case, and if new physics is to rescue this property of the
theory, it must become important at the smallest scale where troubles might be seen.

The momentum scale $\Lambda_{\tilde{k}}$ is of the same order as the threshold energy
$E_{T}$ (assuming the radiating particle is a representative of the lightest species;
if it is not, then $\Lambda_{\tilde{k}}$ is smaller than the typical $E_{T}$).
Assuming that there is indeed a cutoff in the Cerenkov spectrum at a frequency
$\Lambda_{\tilde{k}}$, this will be a lower cutoff than $\omega_{m}$, except in the
limited range of energies $E-E_{T}\ll E_{T}$. At high energies, the continuous energy
loss is no longer given by $P_{<}$, but instead by
\begin{equation}
\label{eq-P}
P=\frac{e^{2}m^{2}}{8\pi}\left(\frac{\theta_{C}^{2}\Lambda_{\tilde{k}}^{2}}
{m^{2}}\right),
\end{equation}
where $\theta_{C}$ means the zero-frequency value of the Cerenkov angle.
Most of the energy is emitted in the highest allowed frequency modes with
$\omega\sim\Lambda_{\tilde{k}}$.
There is no guarantee that any new physics should not be Lorentz violating itself,
and the value of $\Lambda_{\tilde{k}}$ relevant for this calculation may
well depend on the direction of the charge's motion, just as $E_{T}$ depends on
$\hat{v}$. So the expression in parenthesis in (\ref{eq-P}) may depend on
orientation, but its order of magnitude is fixed. It is dimensionless and
${\cal O}(1)$, meaning that the charge radiates at a constant rate, which is
independent not only of the energy $E$ but also of the magnitude of the Lorentz
violation. The smallness of the Lorentz violation is precisely compensated for by
the largeness of the scale at which the cutoff occurs.

It is of course entirely possible that the new physics enters at a scale $\Lambda$
other than $\Lambda_{\tilde{k}}$, though based on naturalness, we would expect this
scale not to be larger than $\Lambda_{\tilde{k}}$. If the true cutoff scale is
$\Lambda$, then (\ref{eq-P})
need only be modified by the substitution $\Lambda_{\tilde{k}}\rightarrow\Lambda$.
In fact, using this formula with $\Lambda=\sqrt{\frac{2}{3}}E$ reproduces the
high-energy form of $P_{<}$, consistent with our earlier interpretation of
$\omega_{m}$ as simply introducing a cutoff at the scale $E$; however, this is
a cutoff that depends on the energy, and hence the power emitted is time
dependent.

With a fixed cutoff, independent of $E$, the $\omega>E-E_{T}$ decay rate $\Gamma$
will also be modified. With a sharp cutoff at a lower frequency, there is simply
no emission of photons this energetic. Until the energy falls low enough that
$E-E_{T}<\Lambda$, $\Gamma$ is zero. A sharp cutoff in frequency is probably
unrealistic, so the rate $\Gamma$ will probably always remain nonzero. However,
the emission of any photon with an energy greater than $\Lambda$ should be strongly
suppressed. The emission of the most energetic photons, which already represents a
slower form of energy loss than the continuous lower-energy emission even in the
absence of $\Lambda$, becomes essentially completely negligible as a loss
mechanism until the charge's energy has fallen low enough so that $E-E_{T}$ is
comparable to $\Lambda$. For $\Lambda=\Lambda_{\tilde{k}}$, this occurs when
$E\sim E_{T}$.

\section{Diffraction}
\label{sec-diff}

Having examined how various ultraviolet cutoffs might come into play,
we shall now discuss a topic that is interesting in its own right
but also serves to demonstrate the complexity with which the multiple scales in the
Lorentz-violating Cerenkov process---particularly the ultraviolet cutoffs---interact.
Diffraction of the Cerenkov radiation turns out to depend crucially on how the
high-energy photon spectrum is cut off. Because it is losing energy, a moving
charge cannot continue emitting Cerenkov radiation forever. The process must have a
limited duration, and
the fact that the resulting track length is finite causes the lower-energy Cerenkov
radiation to diffract. In this way, the recoil from the most energetic photons
indirectly affects the angular distribution of the least energetic ones. The
classical expression for the diffraction width, for the Cerenkov radiation emitted
by a charge that moves superluminally for a distance $L$ without losing
significant energy, is $\Delta\theta\sim
\lambda/(L\sin\theta_{C})$. $L\sin\theta_{C}$ is the distance the charge moves
perpendicular to the direction of the photon emission, assuming that $\theta_{C}$
is not so small as to be comparable to $\Delta\theta$.

However, in the case of interest here, $\theta_{C}$ is always small. Normally,
$\theta_{C}$ gives the width of the cone into which the radiation is emitted.
However, if the diffraction width is comparable to or larger than $\theta_{C}$,
it is $\Delta\theta$ that sets the size of this cone. Of course, the two radiation
cones are structurally very different. In the idealized case of no dispersion and no
recoil, all the photons are emitted on the surface of the cone. When the cone
width is set by diffraction, the photons are smeared out over the full characteristic
angular width $\Delta\theta$. The distance that the charge moves perpendicular
to the direction of photon emission is always $\sim L\sin\theta_{w}$, where
$\theta_{w}$ is the width of the cone, so in the regime where $\Delta\theta$ is
dominant, $\Delta\theta\sim \lambda/(L\sin\Delta\theta)$, or
$\Delta\theta\sim\sqrt{\lambda/L}$.
(For extremely low frequency photons, whose wavelengths are not small compared to
$L$, there is another regime. There the diffractive effects spread the
radiation out over a broad range of angles $\Delta\theta\sim 1$.)

Obviously, the changeover between the $\theta_{w}\approx\theta_{C}$ regime and the
$\theta_{w}\sim\Delta\theta$ regime occurs at the frequency for which
$\theta_{C}(\omega)\sim1/\sqrt{\omega L}$. If the frequency in question is low enough
that recoil corrections
can be ignored and $\theta_{C}^{2}\approx 2(v+n-2)$, the crossover occurs at
$\omega\sim 1/[(v+n-2)L]$. However, determining the correct value of $L$ is
tricky. Two photons with the same frequency
but emitted at two different points can only
interfere in the far field if the Cerenkov angle $\theta_{C}$ does not change
appreciably in the
time between the two emissions. By an appreciable change, we mean one that is larger
than the instantaneous
angular width into which photons of a fixed frequency are emitted; but this
last width is just $\Delta\theta$, which is comparable to $\theta_{C}$ in the
crossover
situation we are presently considering. Therefore, $L$ must be the distance the
charge needs to travel so that its energy loss will narrow the Cerenkov cone by
an ${\cal O}(1)$ factor. For the low-frequency value of $\theta_{C}$ to be cut in
half, the charge's energy must fall from its initial value of $E_{0}$ to
\begin{equation}
E_{1}=\frac{E_{0}E_{T}}{\sqrt{\frac{3}{4}E_{0}^{2}+\frac{1}{4}E_{T}^{2}}}\sim E_{T}.
\end{equation}
In general, for $\theta_{C}$ to change substantially, the energy must fall to
a value comparable to $E_{T}$. The reason from this behavior is fairly clear. The
velocity, upon which $\theta_{C}$ depends, goes to 1 at high energies, depending less
and less on $E$ as $E$ increases; in this regime, $\theta_{C}$ is completely determined by $n$. Only when $n-1$ and $1-v$ are comparable (that is, near the
Cerenkov threshold) does $\theta_{C}$ depend significantly on $E$.

So $L$ is determined by the rate at which the charge loses energy. If the initial
energy is large, the charge will lose most of its energy as it traverses the distance
$L$. With a fixed cutoff $\Lambda$, this requires a time
$\tau_{\Lambda}\approx \frac{8\pi(E_{0}-E_{T})}{e^{2}\theta_{C}^{2}\Lambda^{2}}$. If
the only cutoff
is provided at $\omega_{m}$ by energy-momentum conservation, the time required for
the energy to fall this low is determined by the behavior of $f^{-1}(x)$. In this
case, an
energetic charge loses energy very quickly initially, but the loss rate decreases
with declining $E$. The charge reaches an energy not too far above $E_{T}$ and
decays relatively slowly after that, according to
$E=E_{T}+\sqrt{3\pi E_{T}^{3}/e^{2}m^{2}t}$.
The characteristic time $\tau_{m}\approx\frac{3\pi E_{T}}{e^{2}m^{2}}$ matches our
earlier estimate of $E/P_{<}$, if we take $E\approx 8E_{T}$ to be the characteristic
energy $E$ of the latter estimate, and this value accords nicely with our
qualitative arguments. The time $\tau_{m}$ is also comparable to $\Gamma^{-1}$ in the
$E\sim E_{T}$ regime, so $\tau_{m}$ represents the only relevant time scale arising
from the cutoff $\omega_{m}$.

In any prolonged Cerenkov energy loss process (that is, one in which none of the
$\omega>E-E_{T}$ photons which would bring all further emission to a sudden halt are
emitted),
the $\omega_{m}$ cutoff will eventually become the dominant one, simply because
the energy must reach a point where $E-E_{T}\ll \Lambda$. So if there is a cutoff
$\Lambda$ distinct from $\omega_{m}$, the length $L$ is determined by the distance
the charge travels in the longer of the two times $\tau_{\Lambda}$ and $\tau_{m}$.
Remembering that $\theta_{C}^{2}\sim \frac{m^{2}}{E_{T}^{2}}$, we see that 
$\tau_{\Lambda}$ and $\tau_{m}$ are comparable if
$\Lambda^{2}\sim E_{0}E_{T}$.
For a cutoff $\Lambda_{\tilde{k}}$ at the natural scale $E_{T}$, $\tau_{\Lambda}$ is
larger unless the initial energy $E_{0}$ is also comparable to $E_{T}$. A smaller
$\Lambda$ only makes $\tau_{\Lambda}$ larger, so in all cases with
$\Lambda\lesssim E_{T}$, the time $\tau_{\Lambda}$ predominates.
More generally,
\begin{equation}
\label{eq-L}
L\sim\frac{E_{T}}{e^{2}m^{2}}\max\left(1,\frac{E_{0}E_{T}}{\Lambda^{2}}\right).
\end{equation}

Assuming that there is indeed a cutoff $\Lambda\lesssim E_{T}$, this means that for
frequencies $\omega\lesssim\frac{e^{2}\Lambda^{2}}{E_{0}}$, the diffraction width
$\Delta\theta\sim\sqrt{\lambda/L}$ is larger than the Cerenkov angle $\theta_{C}$.
The location of the change in regimes is tied critically to the value of the
cutoff.
Diffraction originates from the fact that radiation emitted in a finite region of
space cannot be in a pure momentum eigenstate. Some photons must be emitted along
directions other than those specified by energy-momentum conservation considerations.
This suggests the possibility that diffraction might affect the rates of
energy and momentum loss by the moving charge, the potential complications becoming
most serious for very short track lengths. However, it turns out that this is not
actually a problem in this situation.
Since $\Lambda<E_{0}$, the frequencies for which diffraction is important
are all small compared with the cutoff, and photons emitted away from the angle
$\theta_{C}$ do not significantly affect the rate of energy-momentum loss. This was
a necessary consistency check for all our earlier calculations.

\section{Conclusion}
\label{sec-concl}

Vacuum Cerenkov radiation is a very special feature of Lorentz-violating theories.
In this paper, we have described some further properties of the Cerenkov process in
the presence of a CPT-even form of Lorentz violation. The emission rate and other
physically significant quantities are controlled by high frequency cutoffs---and not
necessarily in obvious ways.

Understanding the backreaction of the emitted radiation on the charge was crucial,
since this is what determines the time evolution of the Cerenkov process.
The progress of the physical process is actually rather subtle, and it depends on the
ultraviolet structure of the theory. Photon emissions can be divided into two very
different types, based on the frequencies of the photons involved. Photons with
frequencies below $E-E_{T}$ are emitted more or less continuously, but for more
energetic photons, the quantal nature of the emission is of paramount importance.
As soon as one of these extremely energetic photons is emitted, the charge's
energy drops below the Cerenkov threshold, and the emission process abruptly
terminates. Well above threshold, the total power emitted in the lower-energy modes
is proportional to the square of the ultraviolet cutoff, while the rate at which
$\omega>E-E_{T}$ photons are emitted is increasingly suppressed at higher energies.

If there are no modifications of the
photon sector other than the $\tilde{k}^{\mu\nu}$, the high-energy cutoff for
the photon spectrum arises from energy conservation. The total emission rate for
the lower-energy photons is approximately $P_{<}$, proportional to $E^{2}$. The
decay rate $\Gamma$ describing the higher-energy part of the spectrum is suppressed
at high energies by $E^{-1}$. The charge will lose energy very quickly to start with,
and it is unlikely to decay discontinuously before the energy
has dropped to the scale $E\sim E_{T}$. Once it reaches that regime, the energy loss
rates for the low and high frequency parts of the spectrum become comparable. Most
of the time, the particle will lose an ${\cal O}(1)$ fraction of its remaining
energy above threshold, then terminate the process with a single photon that
disperses all the rest of the energy.

If there is another cutoff $\Lambda$ dictated by new physics, the situation is
different. Naturalness of the quantum corrections to this theory suggest that
$\Lambda$ should probably be no larger than $\Lambda_{{\tilde k}}\sim E_{T}$.
However, as long as $\Lambda<E-E_{T}$, $\Lambda$ is the ultraviolet cutoff that
controls the rate of energy loss. In this regime, the loss rate is independent of
$E$, proportional instead to $\Lambda^{2}$. Moreover, if $\Lambda\sim
\Lambda_{{\tilde k}}$, the energy loss rate is in fact independent of the scale of
the Lorentz violation coefficients $\tilde{k}^{\mu\nu}$ and is simply $P\sim
\frac{e^{2}m^{2}}{8\pi}$. In the regime where $\Lambda$ is the predominant cutoff,
emission of $\omega>E-E_{T}$ photons is all but impossible, since these frequencies
are above the cutoff scale; the discontinuous component of the energy loss is, if
not completely vanishing, strongly suppressed. A charge beginning with a very large
energy will radiate at a constant rate until it leaves the $\Lambda$-dominated
regime. Once $E-E_{T}<\Lambda$, the process is cut off primarily by energy
conservation effects, and the last phase of the process resembles what would be seen
if $\Lambda$ were not present.

The total track length depends on the cutoff, as given by (\ref{eq-L}). Although
this expression for $L$ was not explicitly derived as the total track length, it
does represent the scale of that quantity. The total track length is determined
by how long it takes for the charge to emit just one $\omega>E-E_{T}$ photon. At
high energies, the rate for such emission is very small; the charge must lose
energy until $E\sim E_{T}$ before the rate becomes appreciable. After that, the time
scale required for such a decay is roughly $\tau_{m}$, so the total time (and hence
total track length) is again set by the maximum of $\tau_{\Lambda}$ and $\tau_{m}$.

The strong dependences on how the spectrum is cut off at high frequencies derive from
the fact that, when all effects that might lead to a cutoff are neglected, the
power spectrum grows rapidly at high frequencies. Once a cutoff is included, it sets
the overall rate of energy loss, which determines how the process evolves. When there
are several competing effects that all could potentially cut off the emission,
whichever cutoff is smallest at a given time predominates. At high enough energies,
the energy-momentum cutoff $\omega_{m}\approx E$ will be greater than any
$E$-independent cutoff. So only if there is no other fixed cutoff will $\omega_{m}$
control the emission from the most energetic charges. That there should be no other
energy-independent $\Lambda$ is disfavored by naturalness and causality requirements,
which suggest that new physics counteracting the effects of $\tilde{k}^{\mu\nu}$
should enter at a scale $\Lambda_{\tilde{k}}$ or lower.

The other components of $k_{F}^{\mu\nu\rho\sigma}$ besides those contained in
$\tilde{k}^{\mu\nu}$ are not mixed with any renormalizable coefficients in the
charged matter sector, so naturalness does not dictate any scale at which their
effects are likely to be modified. If all the components of
$k_{F}^{\mu\nu\rho\sigma}$ are of the same order of magnitude, we might expect them
all to be replaced by new physics at the same scale $\Lambda_{\tilde{k}}\sim
m\left|k_{F}\right|^{-1/2}$; however, this is by no means guaranteed. Moreover,
if all the components of $k_{F}^{\mu\nu\rho\sigma}$ are comparable, then
the physical $\tilde{k}^{\mu\nu}$ are constrained by the experimental bounds on the
other components, which can be measured much more accurately because they lead to
photon birefringence.

The $k_{F}^{\mu\nu\rho\sigma}$ coefficients are unique in the photon sector of the
SME, in that they
are gauge invariant and dimensionless. Other forms of Lorentz violation are
parameterized by dimensional constants, and these can introduce natural cutoff
scales on their own. The cutoff dependences may not be so critical as
they are in the case of the $k_{F}^{\mu\nu\rho\sigma}$ terms, but the Cerenkov
processes in the
presence of these other forms of Lorentz violation (including nonrenormalizable
forms) are still quite interesting, and more work is needed to understand them
completely.

\section*{Acknowledgments}
The author is grateful to V. A. Kosteleck\'{y} for helpful discussions.

\end{document}